\documentclass[sigplan, nonacm, 10pt]{acmart}
\setcopyright{none}

\AtBeginDocument{
  }

\usepackage{booktabs}
\usepackage{subcaption}
\usepackage{caption}
\usepackage{xspace}

\newcommand{\sys}{CvxCluster\xspace}

\begin{document}
\date{}

\title{\sys: Solving Large, Complex, Granular Resource Allocation Problems 100--1000x Faster}

\author{Obi Nnorom Jr.}
\email{obdk@stanford.edu}
\affiliation{
  \institution{Stanford University}
  \city{Stanford}
  \state{California}
  \country{USA}
}

\author{Stephen Boyd}
\email{boyd@stanford.edu}
\affiliation{
  \institution{Stanford University}
  \city{Stanford}
  \state{California}
  \country{USA}
}

\author{Philip Levis}
\email{pal@cs.stanford.edu}
\affiliation{
  \institution{Stanford University}
  \city{Stanford}
  \state{California}
  \country{USA}
}

\begin{abstract}
Cluster resource allocation is a multidimensional search problem that finds the best allocation
of tasks to servers. Because the search space grows exponentially, modern approaches 
frame it as a mixed integer program (MIP) or a complex set of search heuristics. 

This paper proposes using a
different approach: convex optimization, which
has extremely fast solution methods. The
research challenge is devising how to transform 
cluster resource allocation into a convex problem that generates
good placements.

We describe \sys, which allocates cluster resources with a
two-stage algorithm.
The first stage solves a convex relaxation of the placement problem to 
yield a principled set of per-machine resource prices.
The second stage uses these prices to drive a
lightweight greedy procedure to place tasks.
Experimental results with Azure traces find that \sys scales to
100{,}480 servers under proportional workload growth and sustains arrival rates
up to 500{,}000$\times$ the baseline trace. \sys runs 100 to 2{,}500$\times$ faster than a state-of-the-art MIP
solver while remaining within 3\% of the optimal objective. 
\sys can support complex constraints such as job anti-affinity,
machine types, and GPU servers.
The key
insight behind \sys is that reformulating placement
as a continuous rather than discrete problem enables much faster methods that 
find solutions just as good or better than prior heuristics.

\end{abstract}

\maketitle
\pagestyle{plain}

\section{Introduction}

Modern datacenters rely on cluster schedulers to place tasks onto shared machines~\cite{borg}.
Placement decisions affect server utilization, system
throughput, fairness across tenants, task completion times, and system efficiency. 
Schedulers must simultaneously reason about multi-dimensional resource constraints (CPU, memory, accelerators,
bandwidth), workload heterogeneity (short interactive services vs. long-running training tasks), and hard/soft policies
such as anti-affinity, locality, and admission control. As clusters and servers grow, a scheduler must handle
an increasing rate of task arrival, and so scheduling
latency becomes a first-order systems bottleneck. A slow scheduler forces smaller batch sizes, limits the
size of a cluster and increases queueing delay, while poor scheduling decisions increase queueing
delay and leave available resources stranded and unused.

The precise, optimal formulation of cluster scheduling is a multidimensional
knapsack problem~\cite{kellerer2004} where resources (CPU, memory) are dimensions,
servers are bags, tasks are objects, and each object has a weight/value. Unfortunately, this is a 
mixed-integer program (MIP), which is NP-complete in the number of tasks and servers~\cite{garey1979}. 
Therefore, state-of-the-art schedulers rely on heuristics and local-search procedures to
provide good-enough placements~\cite{kubernetes,tetris}. Even with these heuristics, existing algorithms struggle to
make high-quality placement decisions on large clusters. Furthermore, they simplify the
problem from some of the details required in practical deployment; POP~\cite{pop}, for example, assumes
all servers are fungible/equivalent, while in practice tasks need to be able to force
co-located replicas to run on different servers for fault tolerance.

This paper shows that there is another way to formulate cluster scheduling
that outperforms prior approaches on all three metrics: it runs orders of magnitude faster, 
places tasks better (under whatever goodness metric one chooses), and can handle
practical complexities such as task anti-affinity constraints. The key insight is
that while mixed integer programs are NP-complete, their close relatives, convex
programs, are not~\cite{convex_textbook}. The mathematics of why are beyond the scope of this paper, but
moving from a discrete (integer) representation to a continuous one opens up
a deep literature on extremely fast solutions to complex, dynamic problems.

This paper presents \sys, a cluster scheduler that uses convex optimization. Convex
optimization has the requirement that the optimization surface is convex: if a local optimum is reached, it is also the global optimum. The key research
challenge, and contribution of this paper, is transforming cluster scheduling
into a convex problem. Once in convex form, \sys makes high-quality placement decisions in time
linear in the number of tasks and independent of cluster size.
The result is a scheduler that scales extremely well with cluster size. 

\sys transforms cluster scheduling into a convex problem with a
two-stage design that decouples global resource reasoning from discrete placement. 
The first stage relaxes the integer placement problem 
into a continuous one.  The relaxation allows fractional
placements; the solver can split a single task across multiple servers, placing a
fraction of its resource demands on each. While fractional assignments are not valid 
in practice, solving this relaxed problem is fast. The relaxed problem is a linear
program, and linear programs are convex. 

Because any local optimum of a convex problem is also the global optimum, every resource's impact on the objective can be exactly quantified by
a scalar dual variable, or \emph{shadow price}~\cite{convex_textbook}. Shadow prices are assigned per server
shape, not per server: this comes from lumping all the available resources of a given server shape
into one pool. The shadow prices are extracted directly from the solver output and act as scarcity signals. 
When many tasks compete for CPU on a server type, its CPU shadow price rises; when memory is abundant, 
its memory shadow price falls. Shadow prices give
the scheduler a compact, globally consistent summary of supply and demand across the entire cluster.

In the second stage, \sys uses the prices to drive a lightweight greedy scheduler. Each pending task computes a net
utility (value minus resource cost) based on its resource demands and the shadow prices. \sys schedules tasks in
order of net utility. This converts placement decisions into a fast
ranking-and-packing procedure guided by globally consistent signals, rather than hand-tuned scoring functions. 

\sys incorporates additional constraints and placement policies either 
directly in the convex relaxation when they admit a tractable representation, or in the greedy placement logic when
they are discrete, policy-driven, or expensive to encode exactly.  This separation allows \sys to remain both fast and
adaptable as constraints evolve, without pushing complexity into a mixed
integer program or a brittle heuristic codebase.

We evaluate \sys on three metrics: scalability, extensibility, and speed. For scalability, we show that \sys schedules clusters of up to 100{,}480 servers
and sustains arrival rates up to 500{,}000$\times$ the baseline trace rate.
We also show that \sys achieves up to 2{,}500$\times$ faster solves than a state-of-the-art MIP
solver while placing within 3\% of the optimal objective.
For extensibility, we demonstrate that constraints such as anti-affinity translate naturally into the linear program's capacity form,
and additional placement policies can be enforced directly in the greedy stage without slowing the convex solve. For
speed, we show that \sys achieves 38$\times$ higher scheduling throughput than DCM~\cite{dcm} on a real Kubernetes
cluster, enabling larger batches and faster reaction to workload dynamics. Overall, \sys suggests a new paradigm for
resource allocation: use convex optimization to extract globally meaningful resource prices, then use those prices to
drive a simple, fast, and policy-flexible placement algorithm.

This paper makes three contributions:

\begin{enumerate}

  \item A new two-stage scheduling paradigm that uses a convex relaxation to compute per-machine, per-resource shadow
prices (dual variables) that summarize global resource scarcity and guide placement decisions.

  \item A lightweight greedy placement algorithm that converts shadow prices into per-task net utilities, enabling fast
ranking and packing while remaining compatible with discrete constraints and policy logic.

  \item A comprehensive evaluation showing improved scalability (128$\times$ larger clusters, up to 100{,}480 servers),
robustness under higher load (500{,}000$\times$ higher arrival rate), and up to 2{,}500$\times$ reduced scheduling
latency relative to MIP baselines.

\end{enumerate}

\section{Background \& Motivation}
\label{sec:motivation}

Making high-quality placements with general constraints is computationally hard largely because the
discrete search space is enormous. A stream of tasks arrives online, each with resource demands
(e.g., CPU cores, memory, and sometimes accelerators such as GPUs), and the cluster consists of
machines with fixed capacities in these same dimensions~\cite{borg,pop,tetris}. Beyond simple
capacity limits, real workloads introduce constraints that couple tasks and machines---for example,
anti-affinity requirements to avoid co-locating replicas, or locality preferences that restrict
feasible machine choices~\cite{google_placement,condor}. In its most abstract and simple form,
placing tasks with memory and CPU requests on a cluster of servers with memory and CPU resources is a
multi-dimensional knapsack problem: it is NP-complete.

The standard formulation used for optimal placement, e.g.\ used in POP~\cite{pop}, is a
mixed-integer program (MIP), which includes both continuous (RAM, CPU) and boolean (placement)
variables~\cite{nemhauser1988,morpheus,rebalancer}. State-of-the-art MIP solvers typically explore this space using
branch-and-bound~\cite{land1960} and related cutting-plane techniques~\cite{gomory1958}, repeatedly solving relaxations while searching
a combinatorial tree~\cite{lawler1966,morrison2016}. This divide-and-conquer search is effective at small scale, but its cost grows
quickly as cluster size and arrival rate increase, which in turn forces smaller batches and lower
scheduling frequency.

Because optimal placement is NP-complete, prior work explores a variety of heuristics to compute
placement. The key idea behind all of these techniques is that by trading off optimality
for speed, solving the optimization problem can become tractable and a system can produce very
good solutions quickly. Kubernetes uses a modular pipeline with scoring and filtering
stages~\cite{kubernetes}. Tetris adapts bin-packing heuristics to multi-resource
placement~\cite{tetris}. Delay scheduling improves throughput and locality under dynamic
arrivals~\cite{delay_scheduling}. These approaches can be fast, but they make locally optimal
decisions without a global view of resource scarcity across the cluster.

Convex optimization offers a fundamentally different tradeoff. Unlike mixed integer programs, convex problems can be
solved in polynomial time because the optimization landscape is such that any locally
optimal solution is globally optimal. The challenge is that placement is a fundamentally discrete
decision: a task must run on exactly one server, but a convex formulation would split it across many.
If tasks could be split, the problem could be solved far faster.

The central question this paper addresses is whether we can exploit the speed and polynomial-time
guarantees of convex optimization to drive cluster scheduling without sacrificing placement quality.
The next section describes how \sys achieves this through a two-stage design that uses the convex
relaxation solely to extract shadow prices, then feeds those prices into a greedy placement procedure
that produces valid, high-quality integer assignments.

\section{\sys Design}
\label{sec:design}

This section describes the design of \emph{\sys}, a two-stage scheduling framework that 
uses convex optimization to schedule tasks to servers in a cluster. 
The key idea in \sys is to decouple \emph{global} reasoning about scarcity and tradeoffs 
across resources, which requires complex computations, from
\emph{local} placement and policy enforcement of individual tasks. This separation  
makes scheduling time predictable and fast, supports complex constraints and
requirements that prior approaches avoid, and scales extremely well 
in cluster size, task arrival rate, and scheduling frequency. 

\sys achieves this decoupling with a two-stage design. In the first stage,
\sys transforms the problem into a convex form that examines the resources available
and the resources requested to compute resource \emph{shadow prices}. These
shadow prices represent 
how much a unit of a given resource can improve scheduling. In this model, 
contended resources that more important tasks need have a higher shadow price.
In the second phase, \sys uses the prices to drive a simple and fast greedy placement
algorithm. 

Compared to search-based heuristics (combinatorial optimization), \sys 
cuts scheduling latency by orders of magnitude, scales to orders of magnitude
clusters, and can easily handle the task arrival rate of these larger clusters.
At the same time, its shadow prices provide a simple and clear signal of 
resource availability and contention, adapting automatically as cluster load 
and resource pressure change.

\begin{figure*}
  \centering
  \includegraphics[width=\textwidth]{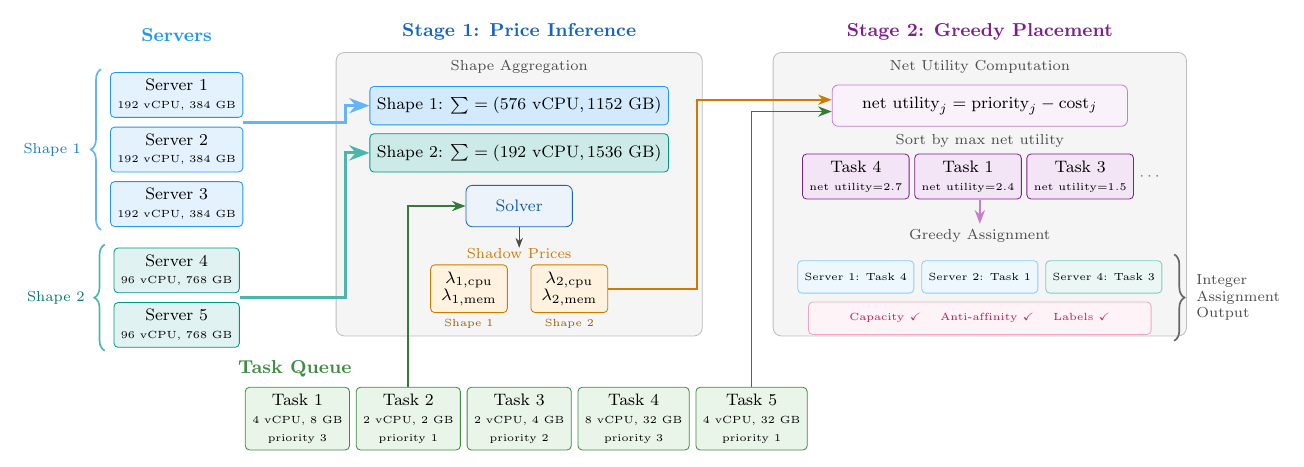}
  \caption{The \sys pipeline. Stage~1 aggregates servers by shape and solves a small linear program relaxation to extract shadow prices. Stage~2 uses these prices to compute per-task net utilities, sorts tasks by utility, and greedily assigns them to servers with fast constraint checks.}
  \label{fig:pipeline}
\end{figure*}

\subsection{Stage 1: Shadow Prices}

Stage 1 infers prices for different resources (e.g., CPU, memory, GPUs) in
the cluster.
Given a standard mixed integer placement 
formulation, \sys relaxes the problem such that tasks could be partially placed,
transforming the mixed integer program into a fully continuous linear program.
This relaxed form is not used to actually place tasks, only to extract the
\emph{shadow prices} of the different resources. 
The shadow price of a resouce quantifies the marginal value, in terms of tasks running,
of adding more of that resource. To allow these computations to scale, 
\sys computes this either per \emph{machine shape} (configuration of RAM,
CPU, GPUs, etc.) to
produce per-shape prices, or globally across the entire cluster to produce a single 
global price vector.

A na\"ive relaxation of a mixed integer program retains its boolean decision
variables, which scale with the number of servers.  \sys’s primary approach to improve
scalability is 
to make its relaxation independent of cluster size by
aggregating servers into a small number of \emph{shapes}. 
A shape corresponds to a hardware configuration: type of processor, number of cores,
amount of RAM, GPUs, etc. For operational simplicitly, any given  cluster in a
datacenter typically has at most tens of shapes.

\subsubsection{Shape aggregation} 
\label{sec:shape-agg}

The algorithm sums the resources of all servers with the 
same shape into a single pool.
With this aggregation, the linear program's decision variables and capacity
constraints are defined over the number of shapes (tens) rather than the number of servers (thousands). 
Since the number of shapes is a small constant in practice, the first stage scales
primarily with the number of tasks to schedule, not with cluster size.
Solving the relaxed set of equations
yields shadow prices associated with server shape. Each server of that shape inherits the global
price vector for that shape. 

\subsubsection{Dual Price Interpretation}
\label{sec:dual-interpret}

\sys relies on shadow prices to summarize cluster scarcity. For example, a high CPU price for a particular machine shape indicates that
CPU is the bottleneck resource on those machines in the current round. These prices adapt dynamically as the workload mix changes.
For example, if arriving tasks become memory-heavy, the price of memory for those machines
will rise, increasing the cost of memory consumption when assigning tasks.

\subsection{Stage 2: Greedy Placement}
\label{sec:greedy}

Computing the shadow prices in stage 1 is the computationally intensive part of scheduling. 
In the second stage, \sys uses the shadow prices to greedily place tasks on servers.
For each task, it computes the \emph{net utility} of running that task on each
machine shape. The net utility is the task's priority minus the cost of resources it
consumes as specified by resource shadow prices.  When a particular resource is scarce, its
price increases, and tasks that consume a large amount of that resource have lower utility. 
They can still be scheduled but will use other resources if they can. For example,
if one machine configuration is requested by many tasks and highly contended, then
other tasks (which don't care) will be placed away from those servers. Similarly, higher
priority tasks will be placed on those limited servers first.

To place tasks, \sys sorts all waiting tasks in descending order of their best net utility across 
all shapes. For each task in order, \sys attempts to place it on the shape where it has the highest 
net utility. It computes the set of candidate servers with that shape that can satisfy the task (have
enough of is requested resources) and places it on a 
server from that set uniformly at random.
If no feasible server exists in a shape, placement falls back to the next-best shape for that task. This procedure is
intentionally simple: the prices do the heavy lifting by creating a near-correct global ordering, while the greedy stage
focuses on fast feasibility and packing.

\subsubsection{Placement Constraints}

Tasks (e.g., in Kubernetes, Borg)
often have placement constraints: they must use a particular processor type or a particular
networking technology. They can also have {\it anti-affinity} constraints, where a collection of
tasks are part of a larger job and for fault tolerance reasons need to be distributed across
different servers.  Placement constraints are enforced by constant-time or low-overhead checks 
during the placement attempt:
\begin{itemize}
  \item \textbf{Capacity feasibility:} ensure the server has remaining CPU/memory/GPU capacity.
  \item \textbf{Machine-specific constraints:} ensure the server satisfies task-required labels/resources.
  \item \textbf{Anti-affinity:} ensure co-location rules are not violated (e.g., no two replicas of the same task on one server).
\end{itemize}

\begin{figure}[t]
  \centering
  \includegraphics[width=\columnwidth]{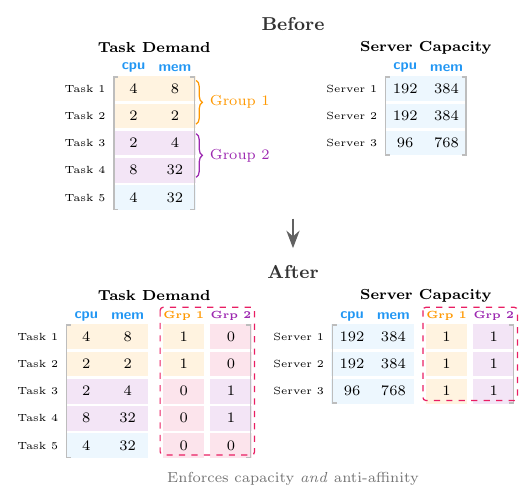}
  \caption{Anti-affinity constraints translate into the capacity form by adding columns to the task demand and server capacity matrices. Each group adds a binary indicator column to the task demand matrix and a limit column to the server capacity matrix, so the existing capacity constraint enforces both capacity and anti-affinity. In this example, each server has a limit of 1 per group, meaning at most one task from each group can be placed on any given server.}
  \label{fig:antiaffinity}
\end{figure}

Each of these constraints can be represented as capacity limits on auxiliary
variables. For example, anti-affinity is represented by dynamically adding per-job variables.
Because only tasks that are part
of that job consider these variables, they do not add
substantial complexity. If a job has a restriction that ``at most 1 task can be on a single server'',
then each task requires one unit of the per-job variable, and each server has 1 unit.
Figure~\ref{fig:antiaffinity} illustrates this translation. With this
encoding, the existing capacity constraint automatically enforces the per-server co-location bound.

Section~\ref{sec:software} describes several engineering optimizations \sys uses to keep
this process fast.

\subsection{Constraint Integration}
\label{sec:constraints}

Constraints must be integrated at placement to ensure that tasks are placed on valid
servers. However, they can also be introduced in stage 1.
Doing so allows \sys to compute shadow prices for these
auxiliary variables, which can allow the greedy algorithm to improve ordering
in placement because the net utility can reflect the constraint's tightness.
The cost is a larger problem (more rows) and an increase in stage 1's
solve time.

Figure~\ref{fig:antiaffinity} shows how this works for anti-affinity constraints.
It illustrates how the solve 
produces additional shadow prices that reflect anti-affinity tightness---without requiring any redesign of the
formulation or the overall pipeline. This translation applies equally to other machine-specific constraints that can be
expressed as per-server capacity limits. Independently, adding or modifying constraints in the greedy stage is localized
to feasibility checks and does not require any changes to the pricing stage.

\subsection{Summary} 

\sys's separation of placement into two stages exposes an explicit tradeoff between problem size/pricing fidelity, 
placement speed, and placement quality under complex constraints. In practice, shape aggregation is the key
scalability mechanism: the relaxation size is independent of cluster size, and in many settings even a single global
cluster-wide price vector is effective for producing high-quality orderings. Meanwhile, the greedy placement remains
intentionally simple---a feature, not a limitation---because the shadow prices provide the global signal that heuristic
pipelines typically approximate with substantial custom logic.

\section{Implementation}
\label{sec:implementation}

This section describes the implementation of \sys, both as
software and the precise formulation of a convex optimization
problem.

\subsection{Software}
\label{sec:software}

\sys comprises two targets that share a common scheduling pipeline: a discrete-event simulator for controlled,
large-scale experiments and a Kubernetes scheduler for real-cluster validation. We implement the simulator in
approximately 4,200 LoC and the Kubernetes scheduler in approximately 3,400 LoC using C++.

The simulator drives scheduling with a priority-queue event loop over three event types: task
arrival, scheduling round, and task completion. At each scheduling round, all pending tasks are batched together and
passed to the scheduling pipeline, which constructs the linear program, extracts dual prices, and runs greedy placement. The simulator
is single-threaded, ensuring deterministic replay of workload traces across methods and seeds.

The Kubernetes scheduler integrates as a custom scheduler that watches for pending
pods and batches them into scheduling rounds at a fixed interval. Each round runs the same linear program and greedy pipeline as the
simulator.

The greedy stage sorts tasks by net utility and, for each task, randomly selects
a feasible server within the best-priced machine shape, since all servers of the
same shape share a shadow price. Anti-affinity constraints are enforced with an
$O(1)$ per-task check that verifies the co-location limit for the task's job
group on the candidate server.

\subsection{Mathematics}

We precisely describe how \sys formulates resource allocation
as a convex optimization problem. We provide these details
because they are the interface to convex solvers: reimplementing
the approach requires the exact representations used.

We define the cluster scheduling problem as a mixed-integer
linear program (MILP).
\begin{table*}
  \centering
  \resizebox{\textwidth}{!}{
  \begin{tabular}{ll rrr rr rrr rrr}
    \toprule
    & & \multicolumn{3}{c}{\textbf{Solve Time (s)}}
    & \multicolumn{2}{c}{\textbf{Speedup}}
    & \multicolumn{3}{c}{\textbf{Objective}}
    & \multicolumn{3}{c}{\textbf{Placement (\%)}} \\
    \cmidrule(lr){3-5} \cmidrule(lr){6-7} \cmidrule(lr){8-10} \cmidrule(lr){11-13}
    \textbf{Servers} & \textbf{Tasks (Jobs)}
    & \textbf{MIP} & \textbf{Shape} & \textbf{Global}
    & \textbf{Shape} & \textbf{Global}
    & \textbf{MIP} & \textbf{Shape} & \textbf{Global}
    & \textbf{MIP} & \textbf{Shape} & \textbf{Global} \\
    \midrule
    25       & 1k    & 0.32  & 0.01  & 0.003 & 29$\times$  & 94$\times$      & 2{,}019  & 2{,}004  & 1{,}989  & 89.6 & 88.1 & 86.6 \\
    250      & 10k   & 58.6  & 0.10  & 0.02  & 586$\times$ & 2{,}566$\times$ & 20{,}457 & 20{,}096 & 20{,}006 & 93.6 & 89.2 & 88.4 \\
    2{,}500  & 100k  & --    & 2.54  & 0.97  & --          & --              & --       & 200{,}874 & 199{,}743 & -- & 89.0 & 88.0 \\
    25{,}000 & 1M    & --    & 134.1 & 89.4  & --          & --              & --       & 2{,}016{,}794 & 2{,}006{,}027 & -- & 89.0 & 88.0 \\
    \midrule
    \multicolumn{13}{l}{\textit{Anti-affinity}} \\
    25       & {\raise.17ex\hbox{$\scriptstyle\sim$}}1k (500)    & 0.32  & 0.01  & 0.004 & 27$\times$  & 85$\times$      & 2{,}077  & 2{,}036  & 2{,}012  & 90.1 & 86.1 & 83.8 \\
    250      & {\raise.17ex\hbox{$\scriptstyle\sim$}}10k (5k)    & 66.8  & 0.12  & 0.05  & 557$\times$ & 1{,}336$\times$ & 22{,}063 & 21{,}565 & 21{,}545 & 90.8 & 85.9 & 85.7 \\
    2{,}500  & {\raise.17ex\hbox{$\scriptstyle\sim$}}100k (50k)  & --    & 3.54  & 3.89  & --          & --              & --       & 214{,}723 & 214{,}450 & -- & 86.3 & 86.1 \\
    25{,}000 & {\raise.17ex\hbox{$\scriptstyle\sim$}}1M (500k)   & --    & 245.0 & 377.1 & --          & --              & --       & 2{,}149{,}762 & 2{,}144{,}453 & -- & 86.6 & 86.1 \\
    \bottomrule
  \end{tabular}
  }
  \caption{Static placement comparison across methods, averaged over 3 seeds.
    Solve time, weighted objective, placement rate, and speedup relative to MIP are shown for MIP,
    \sys (Shape), and \sys (Global).
    \sys achieves up to 2{,}500$\times$ faster solves with comparable objective values.
    MIP achieves higher objective values, with the gap concentrated at the lowest priority level.
    All methods use Gurobi as the solver, with a 600\,s time limit for MIP; a dash (--) indicates the solver did not produce a feasible solution within the time limit.}
  \label{tab:static}
\end{table*}

Let $t$ be the number of tasks to schedule, $s$ the number of servers, and $r$ the number of
resources (e.g., CPU, memory, GPU). The MILP defines the following matrices and vectors:

\begin{itemize}
  \item The task demand matrix $T \in \mathbb{R}^{t \times r}$, where $T_{j,k} \geq 0$ is the demand of task $j$ for resource $k$.
  \item The server capacity matrix $S \in \mathbb{R}^{s \times r}$, where $S_{i,k}$ is the capacity of server $i$ for resource $k$.
  \item The assignment matrix $A \in \{0, 1\}^{s \times t}$, where $A_{i,j} = 1$ if task $j$ is placed on server $i$.
  \item The priority vector $p \in \mathbb{R}^{t}$, where $p_j > 0$ is the priority weight of task $j$.
  \item An all-ones vector of appropriate dimension $\mathbf{1}$ 
\end{itemize}

The program's objective is to maximize the total priority-weighted placement:

\begin{equation}
\max_{A} \;\; p^\top A^\top \mathbf{1}
\label{eq:objective}
\end{equation}

subject to the following 3 constraints:

\textbf{Assignment.} Each task is placed on at most one server:

\begin{equation}
A^\top \mathbf{1} \leq \mathbf{1}
\label{eq:assignment}
\end{equation}

\textbf{Capacity.} Each server's resources must not be exceeded:

\begin{equation}
A T \leq S
\label{eq:capacity}
\end{equation}

\textbf{Anti-affinity.} For each job group $g$ with anti-affinity limit $\ell_g$, the number of co-located tasks on any
server is bounded:

\begin{equation}
\sum_{\substack{j : \, h(j) = g}} A_{i,j} \leq \ell_g \quad \forall\, g,\; \forall\, i
\label{eq:anti-affinity}
\end{equation}

The number of binary variables in this formulation is $t \cdot s$, and the number of constraints grows as
$\mathcal{O}(s \cdot r + t + |\mathcal{J}| \cdot s)$ where $|\mathcal{J}|$ is the number of job groups with
anti-affinity requirements. Even at moderate scale (e.g., $t = 3{,}000$ tasks, $s = 250$ servers, $r = 2$ resources, and no anti-affinity
constraints), the formulation contains 750,000 binary variables; solving this with a state-of-the-art commercial solver
(Gurobi) takes an average of 14.4 seconds across five random seeds. This cost is incurred at every scheduling round, making frequent global
re-optimization impractical under high arrival rates.

\subsection{Convex Optimization}

The MILP formulation in Section~\ref{sec:implementation} contains $t \cdot s$ binary variables, making it intractable at
high scheduling frequency. Relaxing the binary constraint---replacing $A \in \{0,1\}^{s \times t}$ with
$A \in [0,1]^{s \times t}$---yields a linear program that scales as $O(s^2 \cdot t)$ in
practice~\cite{convex_textbook}, where $s$ is the number of servers and $t$ is the number of tasks. After shape
aggregation, $s$ is replaced by $m$, the number of distinct machine shapes, reducing the practical cost to
$O(m^2 \cdot t)$ and making solve time independent of cluster size.

There are two ways to obtain shadow prices from this relaxation. The first is
to formulate the dual of the primal problem (Equation~\ref{eq:objective}), in
which the dual variables correspond directly to the shadow
prices~\cite{convex_textbook}. The second is to solve the primal problem and
extract the dual variables from the solver's output metadata, since modern
solvers compute dual values as a byproduct of the primal solve. At the tested
problem sizes, we use the second approach: solving the primal relaxation
directly and reading the dual values from the solver is simpler to implement
and incurs no significant computational cost.

\section{Evaluation}
\label{sec:evaluation}

Our evaluation answers three questions. First, what is the quality--runtime tradeoff of our optimization core on large
static placement instances? Second, can our scheduler sustain high arrival rates in simulation while preserving strong
performance for high-priority work? Third, how does performance scale jointly with cluster size and arrival intensity?

\subsection{Methodology}
\label{sec:eval-methodology}

All experiments use the Azure Public Dataset V2~\cite{azure_v2}, a 30-day
trace of first-party VM workloads. For the simulation experiments, we construct a cluster of 785 servers drawn uniformly
from five machine shapes spanning 96--192\,vCPUs and 256--768\,GB of memory,\footnote{AWS instance types:
\texttt{m6a.metal}, \texttt{m7a.metal-48xl}, \texttt{c8g.metal-48xl}, \texttt{c6in.metal}, and
\texttt{r8g.metal-24xl}.} totaling 127{,}232\,vCPUs and 450\,TB of memory. We choose 785 servers as the base cluster
size because it is the smallest cluster that can sustain the Azure trace at its original arrival rate without unbounded
queue growth. A smaller cluster lacks sufficient capacity to absorb arriving tasks at the trace's native rate, causing
the pending queue to grow indefinitely; a larger cluster would place tasks trivially and mask scheduling quality
differences between methods. 785 servers represents the hardest scheduling problem, where the cluster
is exactly at capacity.

To explore scheduling scalability, we modify the cluster and Azure trace in two ways: speed and size. 
In speed experiments, we scale the task rate by increasing the arrival rate and proportionally decreasing
task durations. This measures how fast tasks can arrive on a fixed-size cluster and how fine-grained
their durations can become. 
This isolates the scheduler's ability to keep up with arriving work from capacity-driven queueing: any
increase in wait time or queue depth is attributable to the solver running out of computational budget, not to the
cluster running out of resources. 

In size experiments,
we scale the task rate and proportionally increase the cluster size. This measures how large a cluster can
grow with current task sizes.
This reflects the natural growth pattern of production deployments,
where adding capacity accompanies increasing demand. Unlike speed scaling, size scaling measures
whether solve time remains practical as the problem size itself grows---more servers, more tasks per round, and a larger
linear program---rather than whether the scheduler can drain a faster stream of work on fixed hardware.

Finally, in
Section~\ref{sec:eval-k8s-placement}, we deploy \sys on a real Kubernetes cluster to validate that the performance
observed in simulation translates to an end-to-end system, where scheduling decisions must contend with API server
latency, pod startup overhead, and other operational realities absent from the simulator.
For this deployment, we compare against DCM (Declarative Cluster Management)~\cite{dcm}, a system that expresses
scheduling policies as SQL-like constraints and solves them using the CP-SAT solver from Google
OR-Tools~\cite{ortools}. We choose DCM because, unlike POP~\cite{pop} which assumes server fungibility, DCM supports
the same class of placement constraints as \sys, including anti-affinity.

\subsection{Static Comparisons}
\label{sec:eval-static}

To measure how much placement quality the convex relaxation sacrifices compared to an exact solver, 
we use a static scheduling problem.
Static instances---where all tasks and servers are known
upfront---provide a controlled environment to measure placement accuracy in isolation, without confounding factors like arrival
order, queueing delay, or scheduling frequency. Static scheduling also represents the worst-case
problem for \sys: a mixed integer program has complete information and can compute a globally 
optimal result, while \sys's relaxation discards some global information. In a dynamic setting,
both must deal with partial knowledge, as jobs arrive incrementally.

We compare \sys against an optimal mixed-integer programming baseline across
a suite of offline placement instances that scale to up to 1M tasks and 25k servers. Each instance
includes per-task resource demands and per-server capacities, and each task is assigned one of four priority levels
indicating its importance for scheduling. The four priority levels are 1, 2, 4, and 8, and the number of tasks
at each level is inversely proportional, such that the sum of priorities of level 1 tasks is the same as the 
sum of priorities of level 2 tasks.
The objective is to maximize the total \emph{weighted priority} of tasks
placed, subject to capacity feasibility and anti-affinity constraints.

We evaluate two variants of \sys’s pricing stage: \emph{shape
pricing}, which aggregates servers by machine shape and computes a per-shape shadow price vector, and \emph{global
pricing}, which aggregates all servers into a single pooled capacity and computes one global price vector for the entire
cluster.  Both variants use the same greedy placement stage; they differ only in the granularity of the prices used to
compute net utilities.

\subsubsection{Accuracy vs.\ speed.} 

\sys closely tracks the optimal mixed-integer program solution while
substantially reducing solve time (Table~\ref{tab:static}). Global pricing places within approximately 4\% of
the mixed-integer program objective while achieving over 2{,}500$\times$ faster solves. Shape pricing is more
accurate---within approximately 3\% of the mixed-integer program objective---and still provides over
500$\times$ speedup. These results highlight the core tradeoff exposed by the pricing granularity: global prices
yield the smaller problem and the fastest solves, while shape prices recover some lost structure and improve placement
quality.

\subsubsection{Anti-affinity constraints} 

We evaluate \sys under anti-affinity constraints, where each job
comprises multiple replica tasks and at most one replica may be placed on any given server. These constraints couple
tasks within a job and tighten the feasible region, making placement harder. As shown in the lower half of
Table~\ref{tab:static}, \sys continues to scale to 500k jobs ({\raise.17ex\hbox{$\scriptstyle\sim$}}1M tasks) on
25k servers, while MIP fails to produce a feasible solution beyond 5k jobs. Where MIP succeeds (500 and 5k jobs), the
accuracy gap increases modestly compared to the unconstrained setting: shape pricing places within approximately
5\% of the MIP objective and global pricing within approximately 6\%, with speedups of over
550$\times$ and 1{,}300$\times$ respectively. This gap is concentrated at the lowest priority level.
High-priority tasks are placed at near-identical rates across all three methods. \sys’s greedy stage places tasks
in decreasing order of price-adjusted net utility, so higher-priority tasks are placed first and lower-priority tasks
absorb the capacity shortfall. MIP, by contrast, jointly optimizes across all priority tiers and can pack low-priority
replicas more tightly across servers, closing the gap at the bottom of the distribution. In some instances, MIP will
even sacrifice a small number of higher-priority placements when doing so frees capacity for a larger number of
lower-priority tasks, a tradeoff that the greedy stage’s monotone ordering does not exploit.

\begin{figure}[t]
  \centering
  \begin{subfigure}{\columnwidth}
    \centering
    \includegraphics[width=\columnwidth]{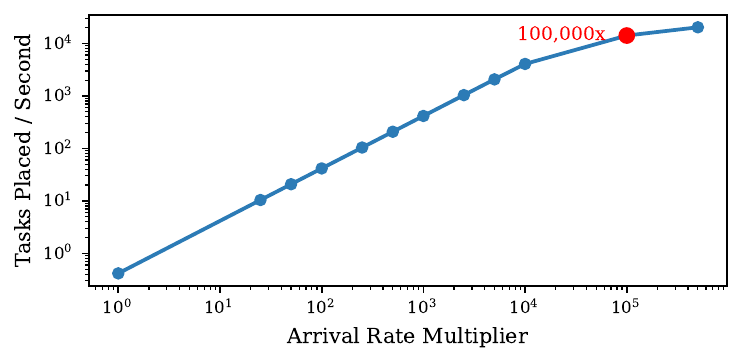}
    \caption{Scheduling throughput scales up to ${\sim}$20{,}000 tasks/sec.}
    \label{fig:scheduling-rate-throughput}
  \end{subfigure}
  \vspace{0.3cm}
  \begin{subfigure}{\columnwidth}
    \centering
    \includegraphics[width=\columnwidth]{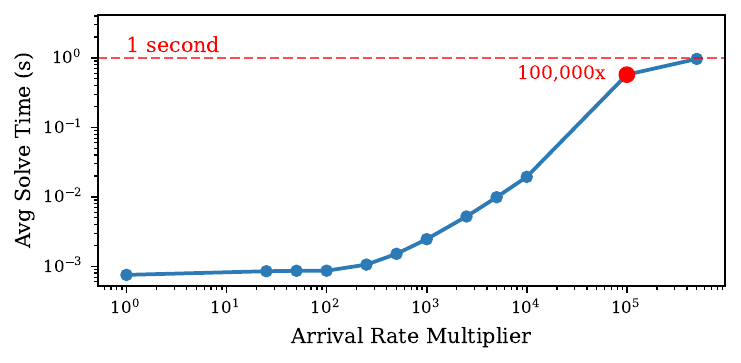}
    \caption{Solve time per round increases from 0.8\,ms at 10 tasks/second
	  to ${\sim}$1\,s at 20,000 tasks/second, at which point the system
	  bottlenecks on the scheduler.}
    \label{fig:scheduling-rate-solvetime}
  \end{subfigure}
  \caption{Scheduler performance under increasing arrival rate multiplier on a 785-server cluster with priority levels.}
  \label{fig:scheduling-rate}
\end{figure}

\begin{figure}[t]
  \centering
  \begin{subfigure}{\columnwidth}
    \centering
    \includegraphics[width=\columnwidth]{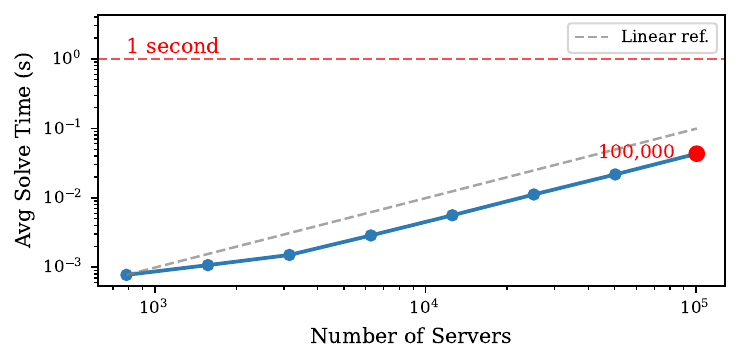}
    \caption{Solve time grows linearly with scale while remaining well under the 1\,s scheduling budget for up a 100,000 server cluster.}
    \label{fig:scaling-solvetime}
  \end{subfigure}
  \vspace{0.3cm}
  \begin{subfigure}{\columnwidth}
    \centering
    \includegraphics[width=\columnwidth]{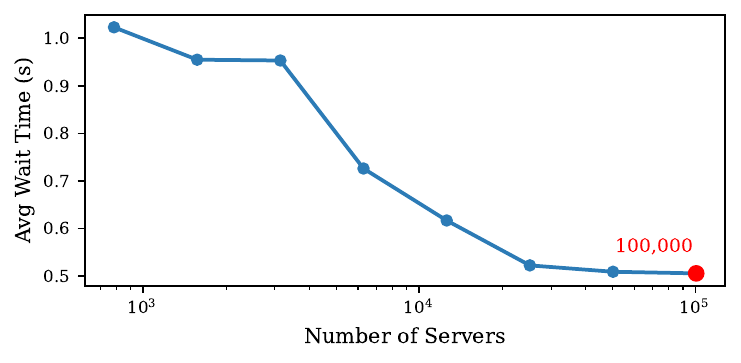}
    \caption{Average wait time decreases from 1.0\,s to 0.5\,s: shape aggregation decouples linear program size from cluster size, so \sys benefits from additional capacity without a proportional increase in solve cost.}
    \label{fig:scaling-waittime}
  \end{subfigure}
  \caption{Scheduler scalability with cluster scaling using a sped-up Azure trace. If scheduling once per
	second, \sys can scale up to a cluster with 100,000 nodes, 127$\times$ the cluster size that
	fits the original trace.}
  \label{fig:scaling}
\end{figure}

\subsection{Speed Scaling}
\label{sec:eval-scheduling-rate}

Section~\ref{sec:eval-static} establishes that \sys places tasks well, but static instances do not
reveal whether the solver is fast enough to run continuously in a live scheduler. In this section, we evaluate \sys
in a simulated online setting, where jobs arrive continuously and the scheduler runs at a fixed 1s
interval. We replay the full 30-day Azure trace on the 785-server cluster
using global pricing to maximize scheduling throughput. The goal is to determine how high the arrival rate can
grow before the scheduler falls behind.

To isolate the impact of arrival intensity from cluster capacity pressure,
we use time scaling, increasing the arrival rate upward while proportionally decreasing job durations, keeping aggregate resource-time
demand approximately constant. This ensures that any degradation in scheduling quality is attributable to the
scheduler's per-round computational budget, not to resource exhaustion. We sweep the arrival-rate multiplier from
1$\times$ to 500{,}000$\times$ the original trace rate.
Figure~\ref{fig:scheduling-rate} shows the results.

\sys remains well within its 1s scheduling budget across a wide range of
multipliers. Per-round solve time is flat at under 1ms up to 100$\times$ and
reaches only 19ms at 10{,}000$\times$, leaving substantial headroom. The scheduler sustains 
up to nearly \emph{100{,}000$\times$} the original trace rate, placing over 19{,}000 tasks/s, at which point average solve
time reaches 574ms with a worst-case of 1.4s. Beyond this regime, at 500{,}000$\times$, average solve
time rises to 966ms with a worst-case of 2.8s, and the task queue grows faster than the scheduler can
drain it.

Average wait time is approximately 1.0s at 1$\times$ and decreases to 0.5s
across the 250--10{,}000$\times$ range. At the original trace rate, jobs retain
their full durations (some lasting
days), so occasional capacity bottlenecks force low-priority tasks to wait many rounds for servers to free up. At higher
multipliers, proportionally shorter durations allow capacity to turn over faster and these bottlenecks disappear. Once
the scheduler approaches saturation, wait time spikes: 7.6s at 100{,}000$\times$ and 13.2s at
500{,}000$\times$, as the queue backlog grows from roughly 50 tasks to 155k and 300k respectively. 
Priority levels preserve differentiation under load: at 500{,}000$\times$, the highest-priority tasks experience an
average wait of 3.0s while the lowest-priority tasks wait 15.7s.
At the 1$\times$ rate, MIP achieves 0.04\,s lower average wait time than \sys, but its average solve
time is 237$\times$ slower. Under dynamic workloads, \sys performs comparably in placement quality
while being fast enough to schedule continuously. As the arrival rate increases, MIP's solve time
exceeds the scheduling interval, making it no longer a feasible scheduling option.

\subsection{Size Scaling}
\label{sec:eval-cluster-scaling}

Production clusters do not stay fixed---they grow as demand grows. This section asks whether \sys's solve time
remains practical as the problem itself scales: more servers, more tasks per round, and a larger problem to solve. We evaluate
\sys's scalability when both the arrival rate and cluster size increase jointly, keeping job durations constant. We
scale from the baseline 785-server cluster up to 128$\times$, reaching 100{,}480 servers.

\sys's per-round solve time grows linearly with scale (Figure~\ref{fig:scaling}): from 0.8\,ms at
1$\times$ to 43\,ms at 128$\times$, remaining well within the 1\,s scheduling interval. 
As cluster and task rates grow, the number of machine shapes doesn't, and so the problem scales
linearly with the number of tasks and not quadratically with their product.
Average wait time decreases from 1.0\,s to 0.5\,s even though the workload grows proportionally
with the cluster. This improvement reflects a well-known property of bin packing: when bins (servers) are large
relative to objects (tasks), almost every object fits into residual capacity with little
stranding~\cite{cxl_pooling}. At 1$\times$, the 785-server cluster is sized to just sustain the trace, so transient
bursts saturate capacity and force low-priority tasks to queue across multiple rounds. At 128$\times$, the
proportionally larger cluster absorbs the same relative bursts with more slack, because each server can accommodate
many tasks and the additional servers multiply the number of feasible packing combinations. Because shape aggregation
makes the linear program size independent of the number of servers, \sys's solve cost scales only with the number of tasks per
round. \sys captures the packing benefits of a larger cluster without any corresponding increase in solver
complexity.

\subsection{Kubernetes Placement}
\label{sec:eval-k8s-placement}

Sections~\ref{sec:eval-scheduling-rate} and~\ref{sec:eval-cluster-scaling} demonstrate strong performance in simulation,
but simulation does not capture the overhead of a real orchestration stack. This section validates that \sys's
throughput advantage holds in an end-to-end Kubernetes deployment, where scheduling decisions must pass through the API
server, contend with network latency, and trigger actual pod lifecycle events.

We deploy \sys on the base cluster of 785 nodes running on AWS using KWOK (Kubernetes WithOut Kubelet), which
simulates node and pod lifecycle events without running real kubelets. Each KWOK node is configured with the same five
machine shapes used throughout the evaluation. The scheduler watches for pending pods and batches them into scheduling
rounds at a configurable interval, running the same linear program and greedy pipeline as the simulator.

We compare against DCM~\cite{dcm}, which batches
pods with a maximum batch size of 50. We run two experiments at 20$\times$ the Azure trace arrival rate: a
one-hour replay with job durations capped at 30\,s, chosen to accommodate DCM's scheduling rate, and a 14-hour
replay to measure \sys's sustained throughput over a longer horizon.

\begin{figure}[t]
  \centering
  \includegraphics[width=\columnwidth]{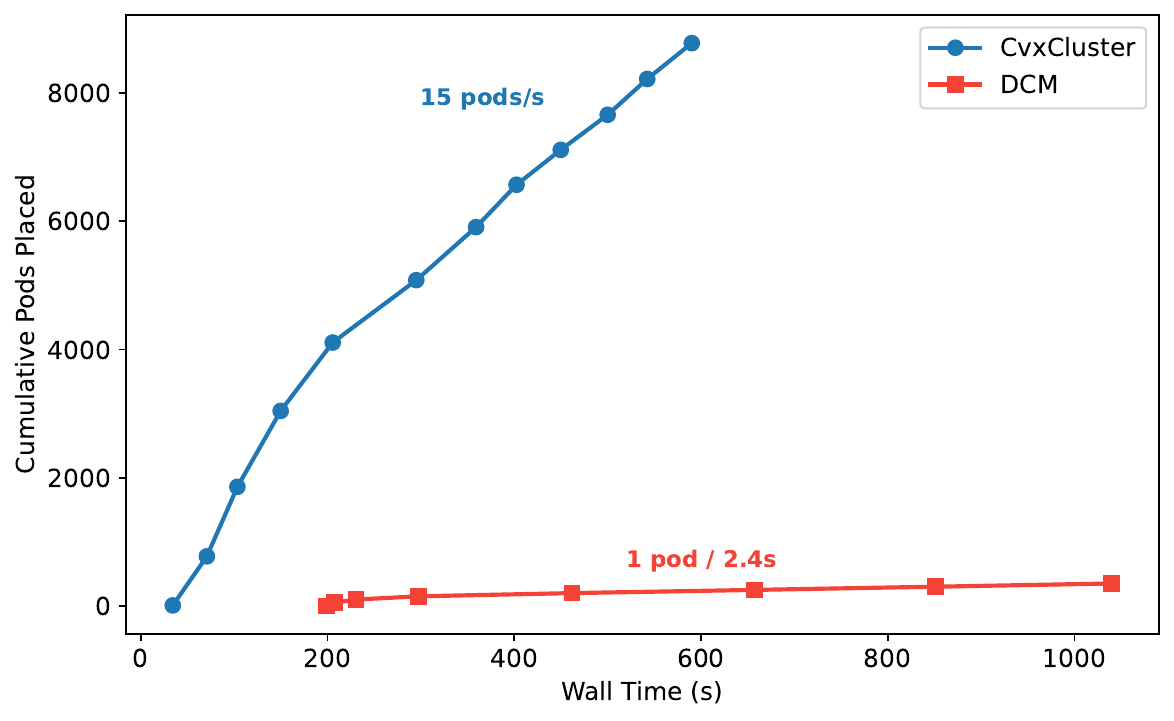}
  \caption{Cumulative pods placed over wall time on a 785-node Kubernetes cluster with 20$\times$ arrival rate scaling.
    \sys places all 7{,}500 pods in under 500\,s while DCM places approximately 400 pods in 1{,}050\,s before
    the experiment was terminated.}
  \label{fig:k8s-placement}
\end{figure}

Figure~\ref{fig:k8s-placement} shows cumulative scheduling throughput over wall-clock time. \sys places all 7,500
pods in under 500 seconds, while DCM places approximately 400 pods over 1,050 seconds---a roughly 38$\times$ throughput
advantage. Even at a batch size of 50, DCM's SAT solver takes longer per round than \sys does with batches 
orders of magnitude larger. Beyond the solver, DCM's end-to-end pipeline includes additional SQL queries and
intermediate processing steps that further limit its scheduling rate. We terminated the DCM experiment early as its
cumulative throughput showed no sign of converging with the offered arrival rate.

\subsection{GPU Extension}
\label{sec:eval-gpu-extension}

Modern clusters increasingly include
accelerators such as GPUs. GPU-equipped servers are expensive and scarce, so the scheduler must avoid wasting them on
tasks that do not need them. This section extends \sys to heterogeneous clusters with both GPU-equipped and
CPU-only servers, verifying that shape pricing handles the additional resource dimension without sacrificing
placement quality or speed. Because the Azure traces do not include explicit GPU requests, we synthesize 
demand by randomly assigning GPU requirements to \emph{50\%} of sampled tasks.

The cluster adds two GPU shapes to the five CPU-only shapes used throughout the 
evaluation,\footnote{GPU shapes: \texttt{g7e.48xlarge} and \texttt{g6.16xlarge}.} each with
distinct CPU, memory, and GPU capacities. Shape pricing produces
separate shadow price vectors for each shape, allowing \sys to distinguish the cost of consuming GPU-equipped capacity
from CPU-only capacity.

We add a penalty term to the objective that increases the cost of placing non-GPU
tasks on GPU-equipped shapes, discouraging them from consuming accelerator capacity unnecessarily. Otherwise 
CPU resources on GPU servers can have equivalent prices to CPU-only servers and CPU jobs schedule on them
interchangeably. The resulting dual prices reflect this preference against placing CPU-only jobs on
GPU servers and are passed to the greedy stage.

We make two final adjustments in the greedy stage to ensure GPU capacity is
reserved for tasks that need it. GPU-demanding tasks are assigned higher priority, increasing their net utility so they
are placed earlier. Separately, GPU-equipped servers receive an additional cost when computing net
utility, making CPU-only servers more attractive.

\sys achieves an identical placement objective to the MIP baseline at 1k and 10k tasks while
solving up to 210$\times$ faster (Table~\ref{tab:gpu-static}). At 100k tasks, MIP runs out of memory while
\sys solves in 17.7\,s. These results confirm that shape pricing extends naturally to heterogeneous clusters: the
additional GPU resource dimension requires no architectural changes to \sys, only the penalty terms.

\begin{table}[t]
  \centering
  \resizebox{\columnwidth}{!}{
  \begin{tabular}{ll rr rr r}
    \toprule
    & & \multicolumn{2}{c}{\textbf{Solve Time (s)}}
    & \multicolumn{2}{c}{\textbf{Objective}}
    & \textbf{Speedup} \\
    \cmidrule(lr){3-4} \cmidrule(lr){5-6} \cmidrule(lr){7-7}
    \textbf{Servers} & \textbf{Tasks}
    & \textbf{MIP} & \textbf{CVX}
    & \textbf{MIP} & \textbf{CVX}
    & \textbf{CVX} \\
    \midrule
    25       & 1k    & 0.24   & 0.02  & 6{,}893   & 6{,}893   & 14$\times$   \\
    250      & 10k   & 42.1   & 0.20  & 69{,}387  & 69{,}387  & 210$\times$  \\
    2{,}500  & 100k  & --     & 4.25  & --        & 688{,}747 & --           \\
    \bottomrule
  \end{tabular}
  }
  \caption{GPU-aware placement comparison, averaged over 3 seeds.
    50\% of tasks are assigned synthetic GPU requirements and placed onto a heterogeneous cluster
    with six shapes.
    \sys achieves identical weighted objective to MIP while solving up to 210$\times$ faster.
    All methods use Gurobi as the solver, with a 600\,s time limit for MIP; a dash (--) indicates
    out-of-memory.}
  \label{tab:gpu-static}
\end{table}

\section{Discussion}
\label{sec:discussion}

\subsection{Placement Efficiency}
\label{sec:disc-efficiency}

\sys's two-stage approach trades a small amount of placement quality for dramatically lower solve times, and the
nature of this tradeoff is predictable. High-priority tasks are placed at near-identical rates to the MIP baseline; the
placement gap is concentrated almost entirely at the lowest priority level. This is a consequence of the greedy stage's
top-down ordering, which guarantees that the most important work is placed first and only best-effort tasks absorb the
capacity shortfall. Across many static solves, MIP often sacrifices placing a few large tasks from the two highest
priority levels in order to pack many smaller tasks at the lowest priority level, achieving better overall utilization
but inverting priority ordering. However, the resulting differences in placement quality are
modest, particularly when weighed against the orders-of-magnitude speedup \sys provides. The objective gap remains
stable at 2--5\% from 1k to 10k tasks, suggesting that the approximation quality does not degrade with problem size.
Within this gap, shape pricing consistently recovers 1--2\% over global pricing by preserving per-shape capacity
structure, at the cost of a larger linear program. This granularity is a tunable knob that operators can adjust based on their
latency budget.

\subsection{Limitations}
\label{sec:disc-limitations}

The linear program relaxation produces fractional assignments that must be rounded to discrete placements by the greedy stage. This
rounding gap is inherent to the two-stage approach and cannot be closed without solving the full integer program. For
clusters where every percent of utilization matters, this loss may not be acceptable.

The scheduler operates in a single-round batch model, placing all queued tasks in one pass per scheduling interval.
If the queue grows large under high arrival rates, solve time increases with the batch size. Conversely, if the interval
is too short, tasks may not accumulate enough for the linear program to produce effective prices. Tuning the scheduling interval
requires balancing these two pressures against the target latency budget.

\subsection{GPU Scheduling}
\label{sec:disc-gpu}

Shape pricing naturally extends to heterogeneous clusters with GPU resources. The evaluation showed that adding GPU as a
resource dimension required no architectural changes to the linear program or greedy stages. The only modification was an additional
cost term in the net utility function that makes GPU-equipped servers more expensive. This changes the dual prices
accordingly, steering placement decisions without altering the core structure of \sys. The same approach should
generalize to other heterogeneous resources such as FPGAs or specialized accelerators.

In the GPU experiments, \sys achieves an identical placement objective to MIP at every scale where MIP produces a
solution. The scenarios we evaluated were heavily resource-constrained, leaving no placement gap between the two methods.
This suggests that when capacity is tight and the feasible region is narrow, the linear program relaxation becomes a closer
approximation to the integer program, and greedy rounding loses little.

\subsection{Potential Improvements}
\label{sec:disc-improvements}

Several directions could improve \sys's placement quality and solve time. First, iterative rounding --- re-solving
the linear program after placing the top-k tasks to update prices with reduced capacity --- could close the rounding gap at the cost
of multiple linear program solves per round. Second, an adaptive scheduling interval that adjusts based on queue depth could balance
latency and batching more effectively than the fixed 1\,s interval used in our experiments. Third, warm-starting the linear program
across consecutive scheduling rounds could reduce solve time significantly, since the problem changes incrementally as
new tasks arrive and running tasks complete. Fourth, finer shape aggregations --- for example, grouping servers by
remaining capacity buckets rather than fixed machine type --- could produce more informative prices without the full cost
of per-server variables.

At larger problem sizes, the greedy placement stage dominates solve time rather than the linear program. At 1M tasks, greedy
rounding accounts for over 90\,s of the 134\,s total solve time for shape pricing. This means there is substantial room
for algorithmic improvements to the greedy stage --- better data structures, parallelism, or approximate nearest-fit
heuristics --- that could significantly reduce end-to-end solve time without changing the linear program formulation.

\subsection{Broader Applicability}
\label{sec:disc-applicability}

The linear-program-plus-greedy framework underlying \sys is not specific to VM scheduling. Any resource allocation problem
with capacity constraints and prioritized demands --- container orchestration, network bandwidth allocation, storage
tiering --- admits the same decomposition: solve a relaxed aggregate problem to obtain prices, then use those prices to
guide fast per-item decisions. The key requirement is that resources are fungible within groups, which is satisfied by a
wide range of infrastructure allocation problems.

Shape aggregation also extends naturally to multi-cluster and federated settings, where each cluster can be treated as a
distinct shape. A global linear program across clusters would produce per-cluster prices that guide placement across geographically
distributed infrastructure, enabling hierarchical scheduling without requiring a single monolithic solver.

The dual prices themselves have economic interpretations beyond scheduling. They represent the marginal cost of consuming
each resource and could be exposed to tenants as cost signals, enabling market-based resource allocation or usage-aware
chargeback systems.

Finally, the prices produced each round could serve as features for learning-based scheduling policies, combining the
theoretical grounding of convex optimization with the adaptability of machine learning. A learned policy could, for
example, adjust the scheduling interval or shape granularity based on observed price volatility.

\section{Related Work}
\label{sec:related}

\sys builds on and connects several lines of work in cluster scheduling: multi-resource bin packing,
optimization-based scheduling, fairness and heterogeneity-aware scheduling, GPU cluster scheduling, and dual-based
decomposition for resource allocation.

{\bf Multi-resource bin packing.} 
Large-scale
cluster managers such as Borg demonstrate the operational importance of efficient placement in production
environments~\cite{borg}.  A large body of research focuses on improving utilization and performance through
packing-aware policies and schedulers, including multi-resource packing approaches such as Tetris~\cite{tetris} and
dependency/structure-aware packing as in Graphene~\cite{graphene}.  These systems motivate our focus on fast, online
placement under multi-resource constraints, but they largely rely on heuristic scoring and rule-based decision logic
rather than explicit global scarcity signals.

{\bf Placement constraints in production.} Real deployments introduce non-capacity constraints---e.g.,
anti-affinity, locality, or machine eligibility---that significantly affect feasible placements and solution quality.
Google’s work on modeling and synthesizing placement constraints highlights the complexity and prevalence of such
policies in production clusters~\cite{google_placement}.  Condor similarly illustrates practical constraint-driven
scheduling in distributed computing systems~\cite{condor}.  \sys is designed so that these constraints can be
enforced cheaply in the placement stage, and optionally priced in the relaxation stage when beneficial.

{\bf Optimization-based and declarative scheduling.} Several systems use mathematical optimization or declarative
formulations to express and solve cluster resource allocation.  POP showed that large-scale granular allocation can be
solved with optimization but is restricted to fungible resources~\cite{pop}.  DCM uses a declarative approach
to  scaling cluster managers, enabling flexibility in expressing policies and constraints~\cite{dcm}.
Rebalancer provides a reusable optimization framework for hyperscale resource allocation at Meta, using an expression
graph and a high-level specification language~\cite{rebalancer}.  These efforts 
highlight the practical tension between expressiveness and solve time as problem sizes and constraint sets grow.
\sys builds on this work with its two-stage design that relies on pricing through a convex relaxation
rather than relying on combinatorial solving.

{\bf Graph-based and distributed schedulers.} 
Sparrow emphasizes distributed, low-latency scheduling through randomized probing and late
binding~\cite{sparrow}.  Firmament and Quincy use flow/graph formulations to incorporate richer global objectives and
constraints~\cite{firmament,quincy}.  \sys is orthogonal to these architectural choices: it can be deployed in
centralized or hierarchical settings, and its Stage~1 pricing can be computed at varying scopes (per shape or globally)
depending on latency and scale requirements.

{\bf Fairness, constraints, and heterogeneity.} Fair resource allocation and heterogeneity-aware placement remain
central concerns in multi-tenant clusters.  DRF provides a widely used foundation for fair sharing across multiple
resource dimensions~\cite{drf}, and Choosy addresses fairness under placement constraints~\cite{choosy}.  
Paragon explores QoS-aware scheduling in heterogeneous datacenters~\cite{paragon}, while Apollo provides a scalable
coordinated scheduling framework for cloud-scale workloads~\cite{apollo}.  \sys focuses on fast placement under
multi-dimensional scarcity; fairness and heterogeneity objectives can be incorporated in the convex stage when convex,
and enforced as policies in the greedy stage when discrete.

{\bf GPU and deep learning cluster scheduling.} Recent systems focus specifically on the demands of deep learning
and GPU clusters, including dynamic adaptation and fairness at scale.  Gandiva, Optimus, and Tiresias address
introspective scheduling, elastic resource allocation, and GPU cluster management for distributed
training~\cite{xiao2018gandiva,peng2018optimus,gu2019tiresias}.  Themis, Shockwave, and Lyra further explore fairness
and efficiency under rapidly changing ML workloads~\cite{themis,shockwave,lyra}, and several measurement and policy
studies characterize heterogeneity and multi-tenant behavior in GPU
clusters~\cite{jeon2019analysis,narayanan2020heterogeneity,chaudhary2020balancing,le2019allox}.  \sys is
complementary: its pricing stage provides a compact global scarcity signal (including for accelerators and
shape-specific resources), and its placement stage can enforce ML-specific constraints and policies without inflating
the convex solve.

{\bf Shadow prices, duality, and decomposition methods.} \sys’s design is inspired by the use of dual
variables and decomposition to expose interpretable marginal signals for constrained resource allocation.  Foundational
optimization texts formalize duality and the interpretation of multipliers as marginal
values~\cite{Ber99,convex_textbook,lagrange}.  Dual decomposition has been used extensively to separate global coupling
constraints from local decisions across domains (e.g., networking, control, and resource
allocation)~\cite{Boy+07,PC06,Chi+07,KMT98,XJB04,Ran09}.  Recent work on price discovery for fungible resources
similarly leverages pricing signals to guide allocation efficiently~\cite{pdm}.  Our contribution is to bring these
ideas to cluster scheduling in a way that makes the convex problem scale independently of cluster size,
letting prices carry the global reasoning while the
greedy stage enforces discrete constraints.

\section{Conclusion}
\label{sec:conclusion}

This paper shows that large-scale cluster resource allocation benefits from relaxing mixed-integer placement
formulations into a convex optimization problem that captures the continuous core of scheduling while avoiding the
combinatorial cost of discrete search. We presented \emph{\sys}, a two-stage scheduler that (i) solves a
reduced-size linear program to infer shadow prices for machine shapes and (ii) uses these prices to drive a fast greedy placement
procedure that enforces task-specific placement constraints.

By making the pricing problem independent of cluster size via shape aggregation, \sys scales predictably as both
arrival rates and cluster sizes increase. In our evaluation, \sys achieves up to 2{,}500$\times$ faster decision-making than
state-of-the-art solver-backed baselines while matching their placement outcomes within 3\%. Finally, \sys
naturally extends to heterogeneous clusters by pricing machine shapes separately, enabling GPU-aware scheduling that
preferentially reserves GPU-equipped machines for GPU-intensive workloads.

\section*{Acknowledgments}

This work is supported by the Building Technologies Office IBUILD Graduate Research Fellowship.

\bibliographystyle{ACM-Reference-Format}
\bibliography{references}

\end{document}